%% file: template.tex
\documentclass{article}

\usepackage{arxiv}

\usepackage[utf8]{inputenc} 
\usepackage[T1]{fontenc}    
\usepackage{hyperref}       
\usepackage{url}            
\usepackage{booktabs}       
\usepackage{amsfonts}       
\usepackage{microtype}      
\usepackage{graphicx}

\usepackage[numbers]{natbib}
\usepackage{doi}
\usepackage{xspace}
\usepackage[most]{tcolorbox}
\usepackage{graphicx}
\usepackage{wasysym}

\usepackage{multirow}
\usepackage[normalem]{ulem}
\useunder{\uline}{\ul}{}

\renewcommand{\sp}[0]{security and privacy\xspace}
\renewcommand{\paragraph}[1]{
\vspace{2mm}
\noindent\textbf{#1}
}
\newcommand{\etal}{\emph{et al.}\xspace}

\usepackage[noorphans,vskip=0.0pt,leftmargin=.1pt]{quoting}

\newcommand{\q}[1]{
\begin{quoting}
    \hspace{1.5pt}
    \noindent\textit{``#1''}
\end{quoting}}

\newtcolorbox{SummaryBox}[1]{enhanced,arc=1mm,outer arc=1mm,
  boxrule=0mm,toprule=0mm,bottomrule=0mm,left=1mm,right=1mm,leftrule=2pt,
  titlerule=0mm,toptitle=0mm,bottomtitle=0mm,top=0mm,
  colframe=blue!50!black,colback=blue!5!white,coltitle=blue!50!black,
  colbacktitle=yellow!50!white,colback=green!5!white,
  title=#1,
  fonttitle=\bfseries\sffamily\normalsize,fontupper=\normalsize\itshape,
}

\title{Trust, Because You Can't Verify:\\
Privacy and Security Hurdles in Education Technology Acquisition Practices}

\author{ 
\textbf{Easton Kelso}\\
\texttt{eakelso@asu.edu}\\
Arizona State University
\and
\textbf{Ananta Soneji}\\
\texttt{asoneji@asu.edu}\\
Arizona State University
\and 
\textbf{Sazzadur Rahaman}\\
\texttt{sazz@cs.arizona.edu}\\
University of Arizona\\
\and
\textbf{Yan Soshitaishvili}\\
\texttt{yans@asu.edu}\\
Arizona State University
\and
\textbf{Rakibul Hasan}\\
\texttt{rakibul.hasan@asu.edu}\\
Arizona State University
}

\renewcommand{\shorttitle}{Education Technology Acquisition Practices}

\hypersetup{
pdftitle={Trust, Because You Can't Verify: Privacy and Security Hurdles in Education Technology Acquisition Practices},
pdfauthor={Easton Kelso, Ananta Soneji, Rakibul Hasan},
pdfkeywords={First keyword, Second keyword, More},
}

\begin{document}
\maketitle

\begin{abstract}
The education technology (EdTech) landscape is expanding rapidly in higher education institutes (HEIs). This growth brings enormous complexity. Protecting the extensive data collected by these tools is crucial for HEIs as data breaches and misuses can have dire \sp consequences on the data subjects, particularly students, who are often compelled to use these tools. 
  This urges an in-depth understanding of HEI and EdTech vendor dynamics, which is largely understudied.
  
  To address this gap, we conducted a semi-structured interview study with $13$ participants who are in EdTech leadership roles at seven HEIs.
  Our study uncovers the EdTech acquisition process in the HEI context, the consideration of \sp issues throughout that process, the pain points of HEI personnel in establishing adequate protection mechanisms in service contracts, and their struggle in holding vendors accountable due to a lack of visibility into their system and power-asymmetry, among other reasons. We discuss certain observations about the status quo and conclude with recommendations for HEIs, researchers, and regulatory bodies to improve the situation. 
\end{abstract}


\input{introduction}

\input{background}
\input{methods}
\input{results-overview}

\input{rq1}
\input{rq2}
\input{rq3}
\input{discussion}
\bibliographystyle{unsrt}
\bibliography{references, references-extra}  

\input{appendix}
\end{document}

%% file: introduction.tex
\section{Introduction}
Digital technologies have become pervasive at higher education institutes (HEIs), integrated into every step of pedagogical and institutional processes\cite{constant-expanding-classroom, russell2018transparencystudentdata}. The huge amounts of data these education technologies (EdTech) collect, encompassing academic records, demographics, behavioral data, mobility and campus life, as well as financial, health, and employment records, have made HEIs lucrative targets for cyber attacks: in 2023, data breaches at HEIs have cost an average of $3M$ USD~\cite{breach-stat}. Simultaneously, reports on the abuse of collected data by service providers and their affiliates for profiling, tracking, advertising, as well as direct selling of the data to data brokers, are on the rise~\cite{student-data-sold, markup-farm-amassing, RemoteLearningSoftwareSellData}. Such practices have deep societal consequences, ranging from privacy violations to potential discrimination for employment and other benefits~\cite{edm_and_privacy, chanensonUncoveringPrivacySecurityK12CHI23, constant-expanding-classroom, russell2018transparencystudentdata}. In the long run, it can also create chilling effects from constant surveillance, undermining students' freedom of expression~\cite{jones2019learning, datafication-HE}. 

Prior research has uncovered privacy and safety issues of EdTech and concerns about those issues from students as well as instructors~\cite{cohney2021virtual, multi-stakeholder-la, edtech-pets22, edtechPETS23}, as they have no choice but to adopt institutionalized technologies~\cite{balash2023educators, compulsory-tech-adoption}. The lack of comprehensiveness in federal data protection laws (such as FERPA~\cite{ferpa}) and their existing loopholes~\cite{zeide2015student, russell2018transparencystudentdata} can be exploited by vendors to evade accountability~\cite{sins-omission, whose-data}. This motivates a need for in-depth scrutiny of the EdTech acquisition practices of HEIs and how they ensure security and privacy during procurement and throughout the life cycle of EdTech, which is largely unexplored. This paper addresses this gap through a semi-structured interview study ($n=13$)  involving people in leadership positions (including Chief Information Security Officers (CISOs), directors, and other senior individuals) at seven HEIs across the US. Specifically, we aim to answer the following research questions:

\begin{enumerate}

\item 
\textbf{Privacy Posture (RQ1):} What regulatory frameworks and internal processes do HEIs use to inform their \sp policy? 

\item 
\textbf{Acquisition (RQ2):} How do HEIs navigate the EdTech acquisition process safeguarding \sp? How are data ownership and liabilities negotiated? 

\item 
\textbf{Post-acquisition (RQ3):} How do HEIs review and perceive data usage, sharing, and accountability concerns throughout the EdTech life cycle?

\end{enumerate}

By exploring the above questions, we provide novel insights in the EdTech acquisition process among HEIs. We highlight our key findings here:

\paragraph{Regulatory leverage: }
The only leverage that HEIs have over EdTech vendors during the acquisition process is the law- and regulation-backed data policies (\S~\ref{privacy-posture-(rq1)}) such as FERPA (Family Educational Rights and Privacy Act~\cite{ferpa}) and HIPAA (Health Insurance Portability and Accountability Act~\cite{hipaa}). 
While data breaches pose significant financial risks for HEIs, vendors often get away with only minor reputational hits (\S~\ref{fed_policies}). 
Additional regulation, where it exists (e.g., at the state level), helps HEIs enforce additional \sp (and as a bonus, accessibility) requirements to EdTech vendors (\S~\ref{state_policies}). 
Though regulations help HEIs, ambiguities surrounding data breach liability limit HEIs' ability to hold vendors accountable (\S~\ref{sharing-liabilities}).

\paragraph{Security concerns:} HEIs have almost no visibility into the security posture and practices of EdTech providers, resigning themselves to rely on questionnaire-based self-reported assessments (e.g., HECVAT~\cite{hecvat}) to evaluate vendor's risk profile and data management policies (\S~\ref{sp_assessment}). 
Few HEIs perform internal audits or penetration tests for the HEI system when EdTech is directly integrated into the system (\S~\ref{refining-HEI-processes}).
Notably, visibility into sub-vendors is even worse, as HEIs lack authority over sub-vendors to undergo assessments, exposing them to potential \sp risks downstream (\S~\ref{sub-vendor-issues}). 
Some small vendors and individual departmental EdTech acquisitions bypass institution-level assessments altogether, leaving HEIs blind and powerless to how their data is used (\S~\ref{sp_assessment}).

\paragraph{Contractual oddities:} Contracts can only go so far in helping HEIs ensure data ownership and protection. 
Data mining, especially relating to AI, is an increasing source of uncertainty and concern in HEI contracts with EdTech vendors. 
While contracts can limit vendors' use of HEI data, the lack of visibility leaves HEIs unaware of data misuse until after the fact (\S~\ref{sharing-liabilities}).
EdTech's off-boarding process is fraught with uncertainty, as HEIs lack the power and means to verify data deletion by vendors.  Thus, even when contracts require the vendors to delete institutional data after their service is discontinued, HIEs cannot verify whether data has been actually deleted; at best they can get a written confirmation (e.g., an email or other document) from the vendors stating that they have deleted all data (\S~\ref{off-boarding-edtech}). We discuss the implications of these findings and conclude with suggestions to improve the status quo in Section~\ref{discussion}.

\paragraph{Contributions.} Overall, our contributions are as follows:
\begin{itemize}
\item We conduct the first qualitative study with \emph{HEI technology leadership} to uncover the EdTech acquisition process and the underlying \sp posture in HEIs that dictates the process.
    \item We provide insights into the nature of EdTech-related institutional policies of HEIs, including how these policies are enforced and monitored during EdTech acquisition and throughout their lifetime.
    \item We highlight regulatory and technical limitations that hinder \sp compliance, especially in the post-EdTech acquisition phase.
\end{itemize}

%% file: background.tex
\section{Background and literature review}

\paragraph{Education technologies.} We use the Wikipedia definition of EdTech that includes any hardware, software, and educational theory~\cite{wiki_edtech}. As such, this definition, when interpreted broadly, encompasses tools that are directly involved in learning activities---such as learning management systems, automated graders, and video conferencing tools---as well as tools that facilitate communications, such as student discussion boards and direct messaging services used in educational settings. That definition also includes algorithms that ship with those tools, such as predictive models (for students' performance or to identify `at-risk' students~\cite{edtech-pets22}), recommending systems for course materials, and automated tutoring agents. For this study, we only consider software tools that HEIs purchase or license from outside vendors and exclude homegrown tools (e.g., research artifacts). 

\paragraph{Data collection, breaches, and liabilities.} 
EdTech now manages nearly every aspect of academic activities~\cite{edtechPETS23, constant-expanding-classroom}, with tech ecosystems constantly evolving with the availability of plug-and-play third-party `apps' (such as app marketplaces for Zoom\footnote{\url{https://marketplace.zoom.us/}} and Canvas\footnote{\url{https://app.learnplatform.com/marketplace}}), paving the way for additional data flows. 
Complexity brings vulnerabilities, as was seen when MOVEit, a popular file transfer tool, was compromised, causing data breaches in almost 900 schools. It exposed students' data including name, IDs, date of birth, contact information, and Social Security number~\cite{Moveit_breach}. We lack an understanding of how HEIs conduct security assessments during procurement, how responsibilities for ensuring secure operations are distributed and maintained throughout the life cycle of EdTech, and how liability is negotiated in case of a privacy or security breach.

\paragraph{Privacy in HEI context.}
Unfortunately, the US does not have any comprehensive data privacy law; the FERPA is the main federal statute that governs the privacy of student data~\cite{russell2018transparencystudentdata}. It requires HEIs to obtain consent before sharing ``educational records'' unless there is a ``legitimate educational interest.'' However, legal scholar Elana Zeide noted that institutions have almost complete authority to define what is ``legitimate educational interest'' and decide what data are protected by FERPA and what are not~\cite{zeide2015student}. Russel~\textit{et al.} noted that the revision to FERPA in 2008 gives more leverage to private companies by including them as ``school officials'' and allows disclosure of data to parties such as contractors, consultants, and volunteers;
and once data enters the marketplace, it can be freely exchanged since FERPA does not apply to data brokers~\cite{russell2018transparencystudentdata}. Such loopholes allow vendors to use vague statements about what data can be collected, used, and for how long, as Paris~\etal reported after reviewing publicly available documents at Rutgers University~\cite{sins-omission}. For example, the contract with Canvas\footnote{\url{https://www.structure.com/canvas}} says that upon contract termination, the university will lose access to the data collected by Canvas, without mentioning if the data will be deleted or the vendor may continue to store and use it~\cite{sins-omission}. This situation also poses legal challenges for HEIs who enroll students from other countries with a stricter privacy policy. As reported in the Findings section, many HEIs we interviewed serve students from the European Union (EU) through online degree programs. Data about these students must be handled following the General Data Protection Regulation (GDPR~\cite{gdpr}), which requires a lawful basis and stronger consenting requirements before sharing data with third parties and imposes restrictions on secondary data usage. Thus, there is a critical need for in-depth scrutiny of HEIs' current EdTech acquisition practices---how they leverage federal and state legislature to guide vendor contracts and negotiations to protect data from breaches and misuse, and what challenges they face. 

%% file: methods.tex
\section{Methods}

To understand the EdTech acquisition practices within HEIs, we conducted $13$ semi-structured interviews~\cite{mack2005qualitative} with individuals in Educational Technology leadership roles in Universities, including Chief Information Security Officers (CISOs), directors, and other senior individuals. 

While our study received exempt status by our Institutional Review Board (IRB), we adhered rigorously to ethical and privacy standards recommended for human subjects studies. This included anonymizing personally identifiable information (PII) such as the participants' and organizations' names, and any other information that could reveal any identity during interview transcription.

\noindent\textbf{Participant recruitment.}  
We recruited EdTech leadership for their pivotal roles in guiding education technology adoption, implementation, and governance at HEIs. Specifically, University leadership provided operational insights, while information security officials offered expertise on security, privacy, and compliance matters. Potential participants were identified through various US institutional websites and the authors' professional contacts. The lead author contacted participants via a recruitment email that outlined study details, compensation, and instructions for interested parties.

We note the significant challenges associated with reaching this target population of leaders at HEIs with specific roles and professional experiences. To address these challenges, we adopted a flexible and iterative recruitment process. We conducted interviews with participants as they became available, continuously inviting new participants. This approach allowed us to engage with a diverse set of participants despite the difficulties in accessing the target population.
To expand the participant pool, we also utilized snowball sampling~\cite{goodman1961snowball}, asking initial contacts to recommend additional participants.

In total, we sent 51 invites and interviewed 13 participants from seven US universities between August 2023 and April 2024.
All our participants had over three years of experience working with EdTechs, with an average experience of $12$ years. 
Working with hundreds of EdTechs, our participants had first-hand experience dealing with vendors and associated challenges HEIs face in the EdTech landscape. 
Regarding educational background, four participants had degrees in business administration, and three had graduate degrees related to EdTech. Although we did not ask our participants directly about their educational background, we noted that information when reaching out to our participants from open sources. Information about the number of EdTechs handled was self-reported during the interview.
More participant demographics are detailed in Table~\ref{tab:participants-info}.

\noindent\textbf{Interview design.}
Our interview questions explored three aspects of the EdTech life cycle, including privacy policies and practices (regulatory compliance), the acquisition process (including criteria for selection and procurement processes), and post-acquisition challenges. We designed our questions to align with participants' specific roles to gain focused and relevant insights. For example, while all participants discussed privacy policies, individuals from learning enterprises provided detailed perspectives on the EdTech acquisition process, while IT security officials offered comprehensive insights into data management and security post-acquisition. This approach enabled us to capture a comprehensive understanding of various perspectives and responsibilities within the EdTech ecosystem, offering a holistic view of how institutions balance educational goals with security and privacy standards. The interview questions for this study can be found in Appendix~\ref{interview_questions}.

\noindent\textit{Pilot study.} The research team collaboratively developed an initial interview questionnaire, which the primary researcher pre-tested through three pilot interviews. 
All interviewees were EdTech users, including one who regularly conducts workshops and develops tutorials on EdTech use. 
They offered complementary perspectives, guiding adjustments in our approach and enhancing the quality of the data collected from the study participants~\cite{thabane2010tutorial}.
These pilot interviews helped clarify questions, improving the language and interview flow (questions such as `factors that HEIs look for when acquiring EdTechs' were expanded with more follow-up questions for greater depth), and revealing potential bias (such as maintaining a neutral tone). After multiple refinement rounds, we finalized the questions and conducted the main study.

\noindent\textbf{Data collection.}
The interviews followed a semi-structured approach, allowing the interviewer to skip or ask follow-up questions as needed, delve into deeper topics, and explore emerging themes.
At the outset of each interview, participants provided verbal consent for their participation, with additional consent requested for audio recording purposes. 
The sequence and emphasis of questions varied based on a) the flow of the conversation and b) the participants' role in EdTech acquisition.
During the interview, participants were provided sufficient time to ask clarifying questions and to gather their thoughts or articulate responses, ensuring their perspectives were accurately captured. 
To minimize interviewer bias, a neutral tone was maintained throughout the process, and the interviewer refrained from expressing personal opinions.
Participants were compensated $\$50$ through an Amazon gift card for a 45-minute online conference call.

Interviews were conducted until data saturation was reached, signifying the point at which no new themes or insights emerge from the data~\cite{fusch2015we}, which is a standard in qualitative research. We collected and analyzed data in parallel and observed that saturation was achieved by the $13^{th}$ interview. 
This sample size is consistent with qualitative research guidelines, which suggest that saturation often occurs within 12-20 interviews~\cite{caine2016local}. 
More details about the analysis process are provided in the next section.

\noindent\textbf{Data analysis}. 
Anonymized transcripts underwent a thematic analysis procedure~\cite{braun2012thematic} using MAXQDA (a qualitative analysis software)~\cite{maxqda}.
To establish an initial codebook, first, two authors independently analyzed four interviews, compared their initial codes, and collaboratively grouped them into broader categories.
The authors used this initial codebook to independently analyze four new transcripts. 
They updated the codebook with new open codes and created a final codebook with $11$ overarching categories by refining already-created categories and resolving any disagreements through discussions~\cite{braun2022conceptual}.
These categories spanned EdTech life cycle at an HEI---from legal guidelines considered during acquisition to factors influencing the acquisition process to security assessments of technologies and post-acquisition challenges.
Authors used this final codebook to analyze three transcripts---two new ones and one of the previously coded transcripts---and achieved an inter-coder reliability score for Cohen’s Kappa ($\kappa > 0.8$)~\cite{mchugh2012interrater}. 
A high inter-rater reliability score, such as this, indicates a consistent coding process, thereby supporting the credibility of the study's findings~\cite{o2020intercoder}
The authors resolved any minor disagreements and continued analyzing the remaining transcripts independently. 
The final codebook can be found in Appendix~\ref{supplememtary_materials}.

Given the qualitative nature of our study, we primarily present our findings qualitatively, occasionally supplementing with counts to highlight prevalent patterns~\cite{mcdonald2019reliability}. 
Additionally, with our semi-structured interviewing approach, it is important to note that any mention of participant counts or IDs associated with specific topics does not necessarily imply exclusivity of those thoughts.

\noindent\textbf{Study limitations}. 
While our study provides valuable insights into EdTech acquisition at HEIs, several limitations warrant consideration. 
First, due to the semi-structured interview approach, not all follow-up questions were asked to every participant, potentially resulting in variations in data depth and comparability across responses.
However, all interviews addressed the key questions necessary to answer our research questions.
Second, although we included participants from a range of universities of varying sizes, our recruitment strategy, sample size, and geographic focus may limit the generalizability of our findings. 
Additionally, generalizability is not typically an ``expected'' attribute in qualitative research due to the specific focus on particular populations and contexts~\cite{leung2015validity}. 
To alleviate this concern, we report our findings to highlight both similar and diverse views from our participants wherever applicable.
Third, biases such as self-reporting and social desirability could have influenced participants' responses. 
We mitigated these risks by framing interview questions neutrally which avoided suggesting any researcher bias and created a comfortable environment for participants to freely share their viewpoints. 

Despite these limitations, this study lays a strong foundation for future research on security and privacy in EdTech acquisition within higher education.
This study identifies gaps in current practices and aims to promote policy and procedural reforms.

\begin{figure*}[ht]
    \centering
    \includegraphics[width=\textwidth]{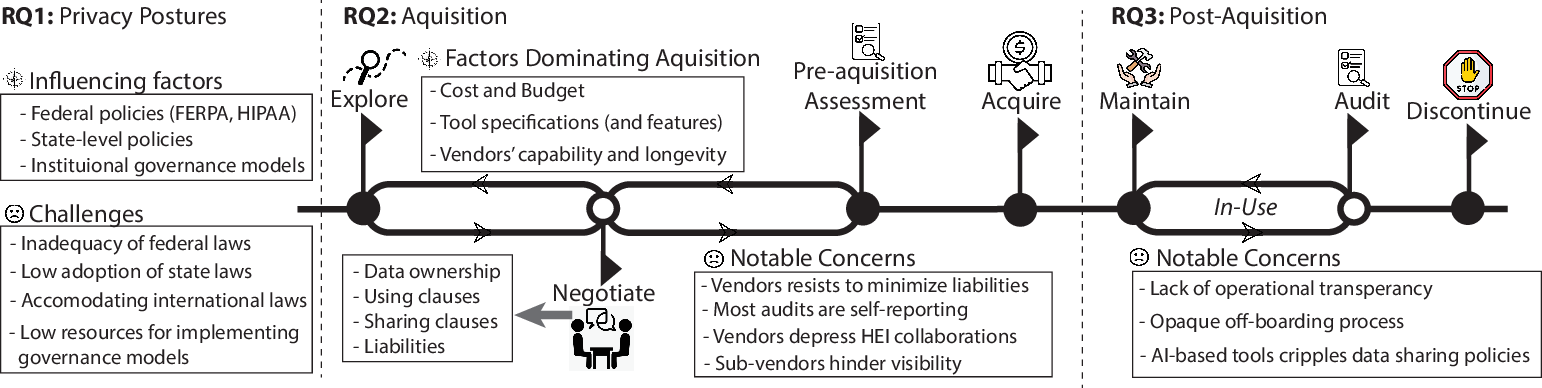}
    \vspace{-10pt}
    \caption{EdTech acquisition and usage timeline learned from our interviews. In light of our final codebook, we also summarize the important findings for each step, both for pre- and post-acquisition phases. Here, \CIRCLE{} indicates a mandatory activity, and \Circle{} indicates an optional one.}
    \label{fig:enter-label}
\end{figure*}

%% file: results-overview.tex
\section{Overview of Results}

Figure~\ref{fig:enter-label} illustrates the EdTech acquisition timeline provided by our participants, key acquisition components and activities, and major findings about each phase in the EdTech acquisition process. Section~\ref{privacy-posture-(rq1)}  details HEIs' privacy posture (RQ1) by examining how federal (~\S~\ref{fed_policies}), state (~\S~\ref{state_policies}), and international regulations (~\S~\ref{international_policies}) inform institutional policies of creating governance models (~\S~\ref{data-governance-model}). Section~\ref{acquisition-phase-(rq2)} examines EdTech acquisition approaches (RQ2, ), covering: a) factors considered in acquisitions (\S~\ref{factors}), b) the role of contracts for data ownership and liabilities (\S~\ref{contracts}), c) pre-acquisition security assessments and their challenges (\S~\ref{sp_assessment}), d) issues with sub-vendors (\S~\ref{sub-vendor-issues}), and e) the need for collaborative groups among HEIs to share vendor experiences (\S~\ref{HEI-collaborations}). Finally, challenges in maintaining, auditing, and discontinuing EdTech (RQ3) are discussed in Section~\ref{post-acquisition-(rq3)}.
These include issues in determining liability in data breaches (\S~\ref{sharing-liabilities}), auditing vendor environments post-acquisition (\S~\ref{visibility}), refining policies and internal auditing processes (\S~\ref{refining-HEI-processes}. We note that while we included contextual information (e.g., mentioned pertinent regulations) and condensed data from several participants, all reported findings were solely derived from the interviews.

%% file: rq1.tex
\section{Privacy Posture (RQ1):}
\label{privacy-posture-(rq1)}
HEIs have privacy policies that, at a high level, guide all institutional activities that may require the collection and processing of data from students and staff, as well as the acquisition and use of EdTech from external vendors. While these policies mostly rely on federal privacy regulations, some HEIs also incorporate state laws or form data governance models with additional terms and conditions.

\subsection{Inadequacy of Federal Level Policies}
\label{fed_policies}
Our interviews revealed that most institutional policies are governed largely by domain-specific federal guidelines, such as FERPA~\cite{ferpa} and HIPAA~\cite{hipaa} for education and health records, respectively. However, most of our participants ($n = 9$), commented that these federal regulations lack adequate protection for sensitive data. For instance, P4 mentioned the need for a comprehensive, rather than sector-specific, privacy law:

\q{One of the larger challenges right now is that the United States doesn't have a federal privacy regulation policy. There isn't a regulatory body that has stepped up into that and there still is debate in Capitol Hill about which regulatory agencies would do that and how they would approach it.}

These existing laws also do not impose sufficiently large penalties on the vendors for violations, as P5 underscored that often vendors may only get a \textit{``little reputational hit''} in case of data misuse and added that:

\q{We need better federal laws... it's a crime that we don't have better protections for people's data.}

In response to these challenges, participants emphasized the necessity for advocacy efforts aimed at addressing deficiencies in federal data privacy legislation.
Some examples include HEIs coming together to establish standards (P2), provide inputs to the legislation (P3), and eventually, create policies that stand the test of time (P5).

\begin{SummaryBox}{Takeaway: Inadequate Federal Level Policies}
    Most HEIs rely on FERPA and HIPAA, meanwhile, vendors may often get away with only a ``little reputational hit'' in case of data breach or misuse.
    Participants advocated addressing these deficiencies in federal data privacy regulations.
\end{SummaryBox}

\subsection{Low Adoption of State Legislation} 
\label{state_policies}
Some states have enacted privacy laws attempting to fill gaps in federal laws~\cite{state-laws}  (such as the Student Online Personal Information Protection Act~\cite{sopipa} in California), but only one institute (out of seven we interviewed) has incorporated state policies into vendor contracts and security assessments.
P11 shared how their institution leveraged state-mandated security questions (different from California) for vendor assessment, which is originally designed for state agencies:

\q{[the state] started looking at all state agencies much more centrally and that created a lot of mandatory questions: what they mean if you answer it no, you're out. So, you have to answer yes. Or you are deemed non-responsive by the state, and you can't then sign the contract. So then we took advantage of that to put some of these [for] security questions, both for security and, quite frankly, for accessibility}

P11's HEI enhanced confidence in data privacy measures by enforcing state-provided five questions---among them, two on security and one on accessibility---during EdTech purchases and noted that ``If you [vendor] don't answer them \emph{yes}, we just say we can't do business with you. So the state helped us there, in a way.''

\begin{SummaryBox}{Takeaway: State Level Policies}
    Most states do not have specific privacy laws, however, the ones that do, allow for their HEIs to leverage laws and achieve higher standards in security.
\end{SummaryBox}


\subsection{Dealing with Intra- and Inter-national Policies}
\label{international_policies}
Several participants discussed the challenges of navigating diverse state and international laws, particularly those with campuses spanning multiple states or enrolling international students.
These institutions will ``follow US law first'' (P5), striving to comply with regulations from other countries such as the European Union’s General Data Protection Regulation (GDPR)~\cite{voigt2017eu}, which imposes huge fines for non-compliance\cite{FGDPR}.
Although P5's institution does not have campuses outside of the US, they discussed challenges towards adopting and prioritizing international laws:

\q{With say GDPR and the Chinese privacy laws (PIPL), you've got some tricky things there. I think if there is a conflict, I don't know how prioritized that would be in terms of meeting the other countries' regulations}

P4, whose institution offers online courses for international students, added \textit{``[HEIs] have to navigate a pretty quickly changing privacy space''}.
To proactively maintain compliance with the rapidly changing (international) privacy landscape, P4's institution adopted the NIST privacy framework~\cite{NIST-PF}, invested in automation and tools like ServiceNow\footnote{A cloud platform used for efficient collaboration and workflow across campuses,~\url{https://www.servicenow.com/solutions/industry/education.html}}, and prioritized privacy protection in their data governance model. We discuss more about governance models in \S~\ref{data-governance-model}.

\begin{SummaryBox}{Takeaway: Intra- and inter-national policies}
HEIs struggle to prioritize and comply with multiple state-level and international policies, often giving precedence to US federal law. However, some institutions take GDPR seriously--largely for huge penalties of non-compliance.
\end{SummaryBox}

\subsection{Institutional Data Governance Model}
\label{data-governance-model}
Five participants (P3, P4, P8, P11, and P12) described how their HEIs have developed privacy-focused data governance models, sometimes referred to as General Counsel, to proactively manage EdTech acquisition-related issues. These models review vendor terms for alignment with HEI privacy standards, control access, and curb abusive practices. P4 discussed the workings of their governance model:

\q{The provost's office has a governance group that meets monthly [to discuss privacy]. That helps [them] form standards and policy and creates a community of practitioners. [They are] sending folks to Privacy Foundation\footnote{The Privacy Foundation mentioned here is a research institution that holds seminars for policy-makers, scholars, and other members of the community to educate people on privacy and policy issues faced by HEIs~\cite{PF}.} specifically to help us build governance models. And then, we spent last year building out a data and privacy governance model for the campus and brought in a really world-renowned expert to help us find how we were going to do data and privacy governance across the campus. A unique model because most people do data governance [only] and they don't have privacy as a subset and we insisted that privacy was more [peer] to the data governments.}

P3 elaborated on their institution's internal policies and procedures, which surpass FERPA and HIPAA requirements:

\q{We have a whole data governance structure in place that governs who can access data, how they can access data, and when they can access data. So, we have a number of policies and procedures in place that really restrict access to only those who have a business need. We audit against that. We review that on a regular basis. So it goes beyond what's just written within the legislation.}

While it was unclear how ``business need'' was defined, and if such needs triumph over data subjects' privacy needs as noted in~\cite{sins-omission, whose-data}. In some institutions, internal policies from the governance body often dominate the contracts during EdTech acquisition. One of the participants (P8) shared:

\q{If a company comes in with a data privacy or data handling agreement that is not aligned with our standards, that will be a stopper for the acquisition}



\begin{SummaryBox}{Takeaway: Institutional data governance policy}
Privacy-focused data governance models enable HEIs to set rigorous standards beyond legal regulations, however, not all HEIs have the ability or staff to achieve this.
Thus, with varied capabilities, HEIs navigate different challenges in advancing privacy initiatives. 
\end{SummaryBox}

%% file: rq2.tex
\section{Acquisition Phase (RQ2):}
\label{acquisition-phase-(rq2)}
This section details EdTech acquisition and how \sp issues influence the procurement process. Several participants outlined the intricate processes involved in procuring EdTech solutions, with P6 summarizing them as follows:

\q{So once we identify that it's [EdTech is] a solution... we think we want to purchase then they [vendors] provide a proposal that outlines what they are offering and what the costs are. We will negotiate that at a product and cost level to say OK this is just enough of these things, we're going to need these features, we're going to need this, whatever. They will put a number against it. Once we have a number that we think is reasonable, we will advance that to procurement. That's where the procurement process requires an audit and what's called a purchasing vehicle.}

This quote reflects broader experiences shared by other participants. For instance, P1, P2, and P3 particularly discussed the necessity of negotiating terms to ensure the inclusion of required features and appropriate pricing.

\subsection{Factors Dominating the Acquisition Process}
\label{factors} 
When asked what factors are considered during acquisition,
participants emphasized the importance of tool features, robustness, and reliability ($n=9$), as well as the longevity of a vendor and its business model (P3, P7, P8). 
For most HEIs, budget is a top priority ($n=9$), as highlighted by P13:

\q{The biggest challenge I have is budgets to be honest. You know, they want the top of the line, but they of course didn't want to spend [the money] on it.}

HEIs also evaluate vendor capabilities, prioritizing those that \textit{``enhance the students' experience''} (P1) and \textit{``[keep] the learning in mind''} (P8), followed by \textit{``how stable the product is''} (P10). Security considerations typically arise later in the acquisition process. On the other end, vendors often prioritize sales over security, as noted by P5:

\q{Most of these companies [are] very excited about the product, they're trying to do good things, but the emphasis is always on sales, and it always feels like security is the afterthought.}

This indicates security is neither a priority by vendors nor by HEIs (at least during the initial stages of acquisition).

\begin{SummaryBox}{Takeaway: Factors impacting EdTech acquisition}
    HEIs prioritize tool features, vendor capabilities, and longevity, but face budget constraints. Security is a secondary concern. Vendors prioritize sales over security.
\end{SummaryBox}

\subsection{Contractual Safeguards and Negotiations} 
\label{contracts}
Contracts serve as crucial tools for HEIs to regulate data collection and usage with vendors.
HEIs critically examine vendor contracts, aiming to negotiate \emph{reasonable} data-use and liability terms and create long-term leverage. These negotiations are often time-demanding.

\noindent
{\bf Data ownership.}
HEIs prioritize maintaining the ownership of the data that EdTechs collect via contracts. According to P2, \textit{``it's always university data''} and P5 noted that their \textit{``legal department pushes back if vendors claim ownership [of HEI data].''} P3 shared a case where a vendor discontinued using student data after \textit{``concerns raised in the higher-ed space around privacy''} and added that \textit{``we're protected via contracts.''} 
As P9 said, \textit{``it all comes down to the agreement.''} 

All our participants' institutions ensured that contracts clearly defined data ownership and the scope of data use. Often, contracts outline ``confidentiality and privacy'' to restrict vendors from ``sell[ing] [institutional] data''. P1 stated:

\q{based on the confidentiality agreement, we would typically not allow them to use our data for any other purposes. It's not to say that you know somebody like [company name] isn't mining the data at a high level to enhance their product capabilities and features. But for the most part, we do not sell our student data and records to be shared by other vendors} 

Vendors, however, can license institutional data for previously agreed-upon uses and are responsible for its secure handling. 
To manage data sharing with vendors, P4 mentioned the use of ``data transfer agreements'' or ``data processing agreements.''
P6 underscored the importance of contractual clarity in licensing data, cautioning against unfavorable attempts of vendors to claim data ownership through ``sloppy agreements.''
Thus, the contractual language is a \emph{pillar stone for HEIs and the only safeguard against vendor issues.}


P3 stated that vendors who understand that ``data that goes into their systems is the property of [university]'' is important. Therefore, according to P3, establishing strong partnerships between HEIs and vendors can be particularly beneficial in the evolving landscape of EdTech. This approach ensures that vendors are not only aware of but also aligned with the needs and expectations of HEIs, facilitating more effective and responsive data management.

\noindent
{\bf Security and privacy terms in contracts.}
Participants (P2, P3, P6, and P11) emphasized the importance of vendors agreeing to institutional terms, especially regarding \sp. P6 highlighted the process of negotiating terms to ensure compliance with institutional policies:

\q{Traditionally, we asked the vendor to agree to [the universities] terms and conditions... part of those terms and conditions includes recognition of security and privacy. And to assure that not only are they meeting them in short term, but that they continue to meet them, and that is a contingent part of our agreement going ahead.}

On the other hand, vendors also push for favorable terms and conditions to reduce their liabilities, as P9 discussed:

\q{If we don't agree with this term that you [a vendor] have in the contract, we want you to take it out. The vendor says, `nope, we need it for our protection, we're not going to take it out' then [the university] needs to make the decision: do we keep that term in there or do we pass and not use the piece of software... it's basically [a] level of risk.}

Lastly, several participants indicated that HEIs struggle to ensure vendors comply with the contracts due to limited insight into vendor operations (\S~\ref{visibility}). P11 expressed how their office of general counsel was working towards finding out how to \textit{``get [checks on vendors] into the contracts, how do we enforce it and how do we check [that they are doing it]?''} 

\noindent
{\bf Contracts under evolving technological landscape.}
The constantly evolving nature of EdTech ecosystems presents further challenges, necessitating continual adjustments to policies and contracts to address evolving threats. P11 said:

\q{The threat surfaces are changing at the speed of light. The bigger the organization, the more you're playing ``catch up.'' We are constantly working to adjust our policies and to adjust our contracts to reflect the current scenario}



Rapid advancements in AI technologies also complicate contractual terms for data management and security. AI technologies require a lot of data for training, sometimes accommodated by extending the definition of ``legitimate educational interest''~\cite{zeide2015student,hooper2022problemswithSchoolData}. P12 shared their apprehensions:

\q{So when we enable AI tools which students want to use, faculty want to use, we want to do research on them. How much of that is actually going back and being used? So is it used to train the model? Where is it stored? What do you do with it? Do you share it with anybody? That's kind of the challenge that I have right now with AI}

\begin{SummaryBox}{Takeaway: Contracts play a critical role!}
    HEIs rely on contracts to define and regulate data use and access by vendors. 
    Conversely, vendors use contracts to minimize their liabilities.
    Data mining by EdTech platforms, especially as it relates to emerging AI technologies, is an evolving contractual grey area of concern to HEIs.

\end{SummaryBox}

\subsection{Pre-acquisition Security Assessments} 
\label{sp_assessment}
During procurement, vendors undergo security assessments, which can take several forms, such as filling out a questionnaire, most commonly created by a third party like Higher Education Community Vendor Assessment Toolkit (HECVAT)~\cite{hecvat}. HECVAT is a 300-question survey that EdTech vendors use to report their security and risk management practices~\cite{Ren-ISAC}. Some HEIs also prepare their questionnaire. Depending on the EdTech's complexity and what data it may collect, HEIs might ask vendors to go through \textit{third-party audits} like Service Organization Control Type 2 (SOC 2), but may also be satisfied by vendors' assessment report. SOC 2 offers HEIs more reliable insights into a vendor's data and security management practices.

\noindent
{\bf Risk-driven security assessment.}
Since security assessments are time-intensive and often involve iterative exchanges, several participants highlighted the importance of determining the risk profile of each tool early in the acquisition process. 
P6 emphasized that their institution's first step is \textit{``to determine the risk profile of a tool''} to tailor the intensity of the subsequent security assessment accordingly. 
P2 explained how different vendors will have different assessments based on the information they are handling:

\q{It's different levels of rigor. A relatively small product with no significant data, we're really focused on assessing security and whether they're going to protect the integration of the service itself and make sure that it's not a pivot place for places that hold data. Then there's a more aggressive one, like we're going to need to see your plans and get your [assessments] outputs, make sure you send them along with your questionnaire report. If you're filling out one of the big assessment reports, send those along so we can keep those on record. } 

Based on the risk profile assessment, HEIs ask vendors to provide audit reports (from either vendors' own assessments or third-party audits like SOC2) or fill up questionnaires like HECVAT. P2 asks vendors ``for their outputs from their own assessments or an independent party’s assessment of their operation to verify that someone beyond them have said that they're doing what they're supposed to.''


\noindent
{\bf Trust in questionnaire-based assessments.}
While popular, the effectiveness of HECVAT and similar assessments was questioned by several participants:  \textit{``sometimes vendors answer in very ambiguous ways''} (P11), and \textit{``honestly, there is probably limited value in the assessment... if they're going to lie to you, you have no way to detect''} (P5).

However, responding to that questionnaire and the response quality can be taken as an indicator of whether a vendor would be a risk to the institution since \textit{``not all vendors feel like they would need to respond to a HECVAT''} (P12). P12 further explained:

\q{If you [a vendor] answered it in two words for every question, I'm not so sure how serious you were when you filled that out... if a vendor really puts in detailed information, then that gives you a little [bit] of a better feeling that... they have the answers that align to what their product is doing. }

The limitations of questionnaire-based assessments were further underscored by the need for manual review of response reports, a task deemed unsustainable with the growing EdTech ecosystems (\textit{``there's too many vendors''} P12). As P11 also noted, \textit{``We really don't have the staff in security to check every single statement that every single vendor makes.''}

Despite these limitations, P12 acknowledged that ``[HECVAT] is useful to gain some trust in that they answered it, but again that's why I like to look at the SOC 2 and the [penetration] test.''
P2, P3, and P8 similarly take extra measures of penetration tests to make sure that \textit{``they're [EdTech] secure''} (P3) and \textit{``we aren't connecting any potential vulnerabilities in our systems''} (P8). 

While this multifaceted approach to security assessment reflects institutions' efforts to mitigate security risks, some EdTech may escape any assessments if they do not go through institutional procurement departments (e.g., being adopted through a less centralized process by the HEI), creating additional vulnerabilities, as P11 disclosed it:

\q{One of the things that we've seen in the last few years is individual departments going out and buying \$40 a month cloud-based applications that never come through the purchasing process. As a result, they’re somewhat invisible to us. ... So we have seen some smaller pieces of cloud software get put in without appropriate oversight. }

\begin{SummaryBox}{Takeaway: Security assessment drivers and trusts}
 EdTech assessment is limited to a questionnaire (commonly, the HECVAT) and, sometimes, a penetration test.
 Most participants expressed concerns about the HECVAT's value, though several suggested that it reflects vendor attitudes and trustworthiness.
 However, even these limited assessment policies can be bypassed by individual departments that make EdTech acquisitions without oversight.
\end{SummaryBox}

\subsection{Sub-vendors and Associated Concerns}
\label{sub-vendor-issues}
HEIs grapple with additional challenges of visibility and the ``unknown'' due to the involvement of sub-vendors (vendors' vendors). This expansion opens the EdTech ecosystem to a larger threat landscape---necessitating HEIs to adapt and adjust their processes accordingly. With more actors playing a part, the risk of unauthorized access to an institution's data increases drastically, eventually requiring the HEIs to not only consider the risk profile of their vendor but also its sub-vendors. Five of our participants (P2, P3, P5, P11, and P12) found sub-vendor breaches concerning as they might lead to an institution's data leak.

\noindent
{\bf Risks and challenges of using sub-vendors.}
P2 expressed that their institution ``struggles'' with sub-vendor issues, describing the situation as ``scary'' due to the sheer \textit{``volume of contracts and reviews.''}
They compared the complexity to navigating ``a spider web.'' P2 elaborated:

 \q{There's a sort of an N plus One problem. You know or like to think about it, if you're familiar with Minesweeper\footnote{\url{https://en.wikipedia.org/wiki/Minesweeper_(video_game)}}, you're going to click on the one file, and it's going to open up a whole new pathway and it's going to be different and unknown in each scenario as opposed to being very well understood.}

In some cases where vendors have extensive subcontracting chains---\textit{``vendors who have sub-vendors who have sub-vendors who have sub-vendors''} (P11)---HEIs need to delve deeply to enhance visibility into their main vendor's sub-vendors. Once vendors (and possibly sub-vendors) send security assessment reports, HEIs' security team may follow up with further questioning. P3 explains:

\q{We will sit down with [the vendor's] security folks [and] go through the HECVAT, get clarity on any kind of questions that they were unable to [answer] on... the HECVAT. }


\noindent
{\bf Additional measures for handling sub-vendors.}
At the start of an agreement, some HEIs require vendors to disclose their vendors' information and mandate audits for sub-vendors based on the initial risk assessment. Some institutions, such as P13's, have dedicated teams working on ensuring that tools from (sub-) vendors that plug into their system are safe by implementing a version of the product on a test bed.



Not all institutions, however, have the resources to thoroughly test (sub-)vendors. P5 expressed that \textit{``smaller universities struggle''} and \textit{``so the process will be as expedient as possible if they've got one.''}
Notably, all institutions stated that they struggle to keep up with the rising number of audits needed to be performed on vendors and sub-vendors. 
P11 speculated on potential solutions to the issue:

 \q{I wonder if this kind of chaos is going to lead back to more internally written rather than vendor written software... So you [the university] have control over that path. I don't know if it will, but I've kind of wondered if we're shooting ourselves in the foot and if we're going to have a little bit of a swing back to on [premise] management or development to help because it is a big problem.}

\begin{SummaryBox}{Takeaway: Sub-vendor management}
    HEIs lack visibility into sub-vendor security practices, forcing them to accept HEIs a large threat landscape that they are unable to fully see into.
\end{SummaryBox}

\subsection{Collaboration among HEIs for Better Leverage}
\label{HEI-collaborations}
The lack of a clear picture of the vendors' \sp posture, exacerbated by reliance on questionnaire-based assessments and limited system access, motivated HEIs to develop collaboration groups, such as Educause~\cite{EDUCAUSE}, Big 10 Academic Alliance~\cite{BIG10}, or the University of California Information Security Council ~\cite{UC}.
These groups, comprising of HEI security officers (CISOs) and other leadership personnel, facilitate the exchange of notes on security practices and experiences working with different vendors.  

HEI collaboration empowers collective influence on vendor practices. While individual institutions may lack leverage, a unified HEI group can effectively advocate for \sp priorities, compelling vendors to take notice and prioritize reforms. P2 describes that power by saying:

\q{The only way that we have found to move [major industry players]
is when higher Education as a whole gets together and says you need to be more responsive to our concerns }

\noindent
{\bf Consultation before finalizing purchase.}
Despite vendors completing third-party audits, five participants (P1, P2, P3, P4, and P5) expressed their practice of consulting their groups before finalizing a purchase. 
This collaborative approach facilitates a quick assessment of vendor relationships, their communication practices, and alignment with the needs of HEIs.
P5 shared their institution's practice of ``bench-marking'' with fellow institutions:

\q{We had a recent issue with a vendor, one of our schools reached out to us and said... ``this is the problem we found [with the vendor]'' and we're like ``okay, thank you! we've got the same problem.'' If I were to acquire something, the first thing we do is [see] what do our other schools [think].}

\noindent\textbf{Information sharing after privacy incidents.}
P3 described how they use their collaboration with other universities to help inform them about privacy incidents and how to make vendors follow necessary obligations:

\q{We had an incident in the recent past [with data leakage] that we, let's just say, leveraged the REN-ISAC that the community is sharing such that others were made aware of some of the issues of this third party. So not only do we leverage the contract and enforce upon the vendor their obligations to notify other institutions that may have been impacted and notify students that may have been impacted, but also leverage that community network that we have to help inform others so that they know the extent of what's happened.}

\noindent
{\bf Factors impeding HEI collaborations.}
While most participants were positive about collaborations, P13 stated how collaborations could be hard due to the differences in priority at differing institutions, saying:

\q{Each university has more of a specific focus on some things, whereas the others don't. Like a lot of ours is like research and medical... There's kind of [a] different focus [at] each [university] and so we at least kind of have our own requirements on IT stuff. So yes, it's different all over the place.}

P5 argues that institutions have limited influence over major vendors and vendors whose primary clientele is outside the higher education sector, and thus, \textit{``[big companies are] probably not going to fill one [a security assessment] out for us ''} (P5).
Moreover, P2 shared that vendors \textit{``purposely isolate''} HEIs to prevent collaboration and comparison between institutions. They shared that vendors strategically avoid discussing terms negotiated with other institutions and instead focus on showcasing what they can offer ``exclusively'' to each university. Vendors further ensure that contractual language prevents HEIs from sharing contract details with their peers. Such tactics make it difficult for institutions to negotiate better terms or leverage their collective power.
\begin{SummaryBox}{Takeaway: HEI collaborations and impediments}
    To help leverage contracts and keep vendor standards, HEIs use collaborations; however, vendors try to isolate HEIs to make it difficult to come together.
\end{SummaryBox}

%% file: rq3.tex
\section{Post-Acquisition (RQ3):}
\label{post-acquisition-(rq3)}
In this section, we discuss the challenges that institutions perceive and experience as they transition to the post-acquisition phase of integrating EdTech into their university networks.

\subsection{Sharing Liabilities during Data Breaches}
\label{sharing-liabilities}
Privacy incidents, such as data breaches or vendors misusing data, are becoming commonplace in the EdTech space~\cite{edtech_breach_article, breach-stat}. P9 identified a large number of EdTech tools per institution---ranging from hundreds to upwards of thousands of services---as one of the major causes of privacy issues:

\q{We deal with so many vendors like probability says someone's going to get [a leak in data]... it's a knotted up ecosystem}

\noindent\textbf{Actions and accountability.}
Regardless of the cause, privacy incidents leave HEIs with little recourse, as P11 stated:

\q{So, it's a violation. What's the remedy? Because how consoled are you if we don't know till after the fact? Your data is already gone.}

Contracts play a vital role in establishing vendors' obligations,
as P5 noted, \textit{``that's [contract is] the most serious leverage we can have.''} 
Similarly, P1 told us that \textit{``if they [vendor] committed to maintaining data privacy, then that's the contractual obligation, and there's not much we can do from that standpoint.''}
In many cases, HEIs make public announcements about the incident, because it is either legally required to give  \textit{``specific notifications''} (P3) based on the data lost or to control reputational damage.  
P2 stated:

\q{there's also just the business part of making sure that our community feels safe and the reputation in the university is known as a place where you can trust to give and provide data and trust that we're going to take care of it}

%

\noindent\textbf{Legal implications and ambiguous liability.}
Privacy incidents can carry significant legal repercussions and hefty fines, particularly if they involve sensitive and protected data (e.g., health or financial data), as discussed by P1:
\q{With HIPAA data you also have to notify the federal government that there was a HIPAA data loss and you could be fined thousands of dollars per record for that loss.}

Determining liability for privacy-violating incidents can be challenging, particularly when considering whether the university or the affected vendor should be held accountable. This challenge is exacerbated if sub-vendors are involved, given that their contract with the primary vendor and the data flow between them is invisible to the university (\S~\ref{sub-vendor-issues}). 
P11 expressed their concern regarding this ambiguity:

\q{Is it the university [who is liable]? Is it the vendor of the university? Is it the sub-vendor or the sub-sub-vendor? Who is liable? To date, the Department of Education and some very, very sketchy advice says ultimately, the university runs the application, they're liable.}

While contracts help to some extent in cases of privacy violations, the need for standard guidelines and accountability mechanisms becomes crucial in navigating these situations.

\begin{SummaryBox}{Takeaway: HEI vulnerability to vendor breaches}
    Vendor breaches pose significant liabilities for HEIs, with contracts often leaving ambiguities and inadequate recourse once data is compromised. Moreover, ambiguities in federal regulations in data breach liability (e.g., as it relates to FERPA) limit HEI's ability to hold vendors accountable.
\end{SummaryBox}

\subsection{Visibility and Transparency Concerns}
\label{visibility}
\noindent
{\bf Lack of auditability.}
The ability to perform or require third-party audits is a large part of the acquisition process, it allows for HEIs to develop a level of trust and transparency with technologies (\S~\ref{sp_assessment}).
However, once the EdTech are acquired, institutions lose their ability to audit those technologies until the next licensing cycle.

HEIs typically maintain contracts ranging from three years (as noted by P1 and P11) to five years (according to P6), aligning with state laws and institutional policies.
Consequently, the maximum duration between audits for an EdTech is dictated by the length of the contract.
P3 highlighted this gap and described their institution's approach to scrutinizing purchases during the acquisition process.

 \q{We don't have an audit process after the fact. I think that's probably a gap... we do as much scrutiny as we can as a part of that purchase process, working with their technical folks to kind of understand their environment. What processes and controls that they have in place. }

\noindent\textbf{Trust issues associated with the obscurity.}
The inability to audit tools in operation or the vendor's environment invokes potential trust issues, and P5 said that they \textit{``don't particularly have a way to test whether a vendor is abusing our data.''}

Participants, including P11, voiced frustrations over vendors' lack of transparency regarding changes, particularly regarding sub-vendors:

\q{what are you [a university] going to do if they change a sub vendor? Because no vendor on the planet is going to say `oh we're changing our database vendor. Don't forget to tell [this university]' that ain't gonna happen.}

When institutions find out about such changes during contract renewals, products must go through the entire auditing process again. P11 highlighted concerns that HEIs remain unaware of until these contract renewals.

\q{We continue to refine the processes, but that is one of the product managers responsibility... [Asking] where's that data going? Is the place that it is going, did they get a security evaluation? When was the last time they got a security evaluation? If somebody changes a vendor, until our contract gets renewed and we redo another evaluation, we'd never know that unless the vendor tells us. There's no contractual trigger.}

\begin{SummaryBox}{Takeaway: Post-acquisition trust and obscurity}
    HEIs struggle to audit vendor environments post-acquisition due to long contract durations, limiting visibility into vendor practices until the next renewal. 
    Trust concerns are raised as HEIs remain unaware of changes in vendor processes or sub-vendors.
    Thus, HEIs adapt to rely on trust rather than active oversight in an environment of obscurity.
\end{SummaryBox}

\subsection{Refining HEIs’ Processes and Practices }
\label{refining-HEI-processes}
Because HEIs cannot audit EdTech in operation, some of them have implemented measures to keep their internal system secure and regularly refine those practices.

\noindent\textbf{Internal audits and penetration tests.}
To prevent external tools from introducing vulnerabilities into HEIs' internal systems, some institutes conduct annual internal audits and penetration tests, especially with vendors that directly plug into their systems. P3 discussed it:

\q{I don't know that we [can] do testing from a security and privacy perspective to ensure that the products are not doing something they shouldn't be doing. The only thing that we may do is a penetration test scenario which we do for the most part on internal systems to make sure that they're secure.}

HEIs rigorously review their practices as a crucial protective measure. P2 highlighted their institution's periodic audit procedures, noting that top systems are reviewed every two to three years and ancillary systems every three to five years. 
P2 elaborated on the details of their internal audit process:

\q{We have an internal policy that does at least two audits on internal systems and services, that does extend [to] the cloud. The audit asks us how we do data security and backup, ask industry standard questions: are we following the right processes, the right controls?}

However, P2 expressed frustrations that such procedures are ``unsustainable for checking every vendor that their institution has.'' To mitigate this, P1 and P3 take proactive steps to strategically reduce the number of technologies employed and opt for solutions that integrate multiple capabilities.


\noindent\textbf{Adapting to evolving technologies.} 
As part of continuously refining audit processes, some HEIs are considering AI tools, while acknowledging associated risks. 
AI could help HEIs protect their own data and increase surveillance over vendor tools. P11 shared their thoughts on AI-based tools:

\q{We see an awful lot of new security tools that are AI based...how do you know the algorithm for that AI is good enough to actually work. Well it works 90\% of the time, well what about that other ten? Our CIO is concerned that AI will help the bad guys penetrate the network. We also see AI coming on the good guys side within the tools to help monitor millions of rows of log data that our people can't get all the way through.}

Overall, HEIs strive to adapt to changing environments and preserve control over their own systems. 
In the whole \textit{``you have to trust them [vendors] at some point''} (P1), innovative strategies become crucial for HEIs to address post-acquisition challenges and ensure data security.

\begin{SummaryBox}{Takeaway: Refining processes and practices}
    While not having the ability to confirm that vendors are performing their obligations, HEIs focus on their own practices and look ahead at the growing landscape of emerging tools such as AI that could assist or harm them.
\end{SummaryBox}


\subsection{Off-boarding EdTech}
\label{off-boarding-edtech}
EdTech discontinuation is usually triggered by price increases (P1, P3, P6, and P8), vendor disputes (P5, P6, and P8), or the identification of better tools (P1, P2, and P8). Regardless of the reason, discontinuation raises an important question: What happens to the data collected by the vendor after the discontinuation?

When services are discontinued, vendors typically initiate an off-boarding process outlined in the contract. This process ideally involves erasing all institutional data from their systems and providing a confirmation document. However, vendors may not be as cooperative as one might like; P2 highlighted that:

\q{Once you tell them [cloud vendors] you're not buying anything from any of them anymore, they are much less inclined to help you off board ... it might be as little as half of the off boardings, we get the confirmation that --- Yes, everything's shut down and all [the university's] data is gone}

While vendors may be contractually obliged to provide a \textit{``written assurance''} (P5) about data erasure, universities need to pursue vendors for that, and that  depends on the sensitivity of the records, as P2 explains:

\q{Unless it's in the contract, unless we want to pursue long squabbles with a disinterested party, we essentially have to come to a conversation. If we had 200,000 of our alumni with all their social security numbers (SSNs) on it, you can bet we're going to pursue it. If we have 200,000 people with their addresses and a phone number, we're likely to pursue it. If we have 1000 people and it's just their first name, we're probably going to write it off.}

P12 emphasized that while a signed assurance provides \textit{``some legal protection''}, it may not fully address concerns if data remains with the vendor despite claims of erasure. P5 shared that due to vendors' non-cooperation, the off-boarding for complex systems could take multiple years. P3 highlighted a trend towards centralized systems managing various functions from \textit{``student information systems to HR to finance.''} They expressed concerns if institutions wish to adopt such systems:

\q{Once you get there, it's going to be harder to get off and go anywhere else because you've got one integrated system. There are advantages to that, but then you're locking yourself in and it gets harder to break yourself away.}

\begin{SummaryBox}{Takeaway: Off-boarding process}
    When discontinuing an EdTech, HEIs have no power or ability to investigate if their data has been fully erased from a vendor's system. HEIs are made to trust a written verification made by the vendor itself, and even that is only sometimes provided.
\end{SummaryBox}


%% file: discussion.tex
\section{Discussions and conclusions}\label{discussion}
This study aims to surface specific \sp challenges faced by HEIs that come from the specific tools they use, the resources and processes they have in place, and HEI-specific policies. Thus, while some findings may apply to many HEIs, we do not aim for generalizability. 
Below, we discuss our results by highlighting key challenges and providing recommendations based on our study that can help strengthen \sp within HEIs regarding EdTech acquisition, retention, and discontinuation.

\noindent\textbf{Limitations of FERPA in safeguarding HEI.}
Overall, our study suggests that HEIs' privacy posture is largely underlined by the protections offered by FERPA (\S~\ref{fed_policies}).
This result is consistent with Brown and Klein~\cite{whose-data}, who found that out of $151$ institutional policy documents, $81$ were FERPA-related. 
Given the limited data types that are protected by FERPA~\cite{ferpa}, the coverage of entities (only federally funded organizations)~\cite{ferpa}, and the existence of loopholes that can be leveraged to evade compliance altogether~\cite{loopholes}, the privacy of institutional data is in jeopardy.

\noindent\textbf{Security assessments dilemmas.} Although compliance with FERPA and HEIs' security practices is mandated~\cite{ferpa}, currently there is no way to guarantee it from the vendors' end (\S~\ref{sp_assessment}). 
HEIs mostly rely on a questionnaire-based assessment of EdTech before procurement (e.g., HECVAT~\cite{hecvat}), which does not provide any actual visibility into EdTech and its auditing capabilities. 
Some HEIs even lack resources to examine those assessment reports in the face of a growing number of vendors (\S~\ref{sub-vendor-issues}). 
We note that assessments like HECVAT and SOC2 focus on securing the data from unauthorized access, but they do not cover how this data can be used by ``authorized'' entities.
Alarmingly, many vendors are unwilling to complete even these basic assessment procedures, highlighting a critical gap in data protection that has significant implications for student privacy and institutional accountability.

\noindent\textbf{Contractual challenges and power asymmetry.} Contracts dictate data ownership and use. 
However, we uncovered that HEIs face challenges to create contracts with concrete and privacy-focused terms and conditions regarding data collection and use (\S~\ref{contracts}). 
First, the web of sub-vendors (\S~\ref{sub-vendor-issues}) exacerbates the issues of ensuring robust data protection practices at the contract level.
Second, even when HEIs try to address ambiguity in security assessments and contract documents or revise terms to reflect proper data use and accountability, vendors \emph{often} push back. 
This reveals a significant power asymmetry between vendors and HEIs, including those among the largest public universities in the US.
While some universities form coalitions to improve their negotiating position, these efforts face collaboration challenges and often exclude smaller universities and community colleges.
Third, HEIs face inherent difficulties in making concrete statements about data collection since they cannot independently examine the tools, making it incredibly difficult to verify from outside what these tools are doing, what data they are collecting, and where they are sending this data. 
Consequently, this power imbalance and lack of transparency forces HEIs to trust vendors' assurances, a practice fraught with risk, as highlighted by our participants.


\noindent\textbf{Data ownership challenges.} Having addressed the broader contractual challenges, our study also uncovered grim realities regarding data ownership and use in the context of EdTech acquisitions (\S~\ref{contracts}). 
Firstly, HEIs strive to retain ownership of their data, but achieving exclusive control often depends on the institution's negotiating power.
Secondly, contracts may permit vendors to use HEI data without owning it, raising concerns about control and usage rights.
Thirdly, vendors can grant similar data usage privileges to sub-vendors, who can change at any time and may have arbitrary contracts that allow them to share or sell data with other parties (including the primary vendor) or data brokers without any restriction.
Lastly, vendors might gain sole ownership of data if services are discontinued, as HEIs lack mechanisms to ensure data deletion.
The complex relationships of HEIs with vendors and sub-vendors underscore the necessity for further research to unravel the implications of these dynamics on data control and privacy in EdTech acquisitions.


\noindent\textbf{Obscurity of (improper) data utilization.} Data also creates new data. For example, behaviors often correlate with demographics and other personal factors, which can be ``inferred'' from interaction data collected by various EdTech tools~\cite{edtech-pets22}, allowing vendors to learn more about the students. 
In addition to deciding the ownership for this newly created data, their use has profound ethical implications~\cite{kyritsi2019pursuit} and the potential to disproportionately harm vulnerable groups (e.g., outing a non-binary student by ``inferring'' their gender~\cite{edtech-pets22}). 

Indeed, one of the primary uses of student data is creating such machine learning-based predictive models~\cite{edtech-pets22}. 
Companies like Instructure, which developed Canvas, offer services based on these models~\cite{canvas-la}. 
When such companies are acquired by for-profit equity firms, as seen with Canvas~\cite{canvas-data}, it further intensifies the ethical issues surrounding the digitization of education. 
The ethical standards of the previous company may not necessarily be upheld under new ownership, and the prospect of misuse or sale of extensive student data raises serious concerns~\cite{loopholes, student-data-sold}.
Educational technologists are increasingly raising alarms about such company takeovers and other issues with the EdTech ecosystem~\cite{datafication-HE, jonesMatterTrust2020}. 




\noindent\textbf{Risks and liabilities of AI/ML in EdTech landscape.} We also learned that while (sub-)vendors profit from data by building and selling new (AI/ML-based) services, they don't share the associated risks and liabilities (\S~\ref{refining-HEI-processes}). 
Moreover, vendors often include terms that waive their liability, as also noted in prior research~\cite{sins-omission}.
Even when liability terms are included in contracts, the complex chain of (sub-)vendors---who can update their privacy policies without HEIs' knowledge---makes it nearly impossible to assign liability in the event of a privacy violation. 
Although HEIs may seek protection for the data collected by vendors, there are no safeguards for the ML models trained from that data.
Notably, recent research shows that ML models are susceptible to attacks that can reveal the very data used to train them through model inversion or querying~\cite{privacy-attacks-ml-survey}.
These emerging concerns necessitate a reevaluation and enhancement of regulatory frameworks.

\noindent\textbf{Recommendations:} 

\noindent\textit{\textbf{For regulators.}} We strongly advocate for comprehensive federal privacy laws to provide HEIs with baseline protection regardless of their negotiating power with vendors. 
We also recommend federal or state mandates for standard contracts with default data protection clauses that will apply to all (sub-) vendors to prevent privacy issues caused by complex vendor chains and non-compliant EdTech deployments.


\noindent\textit{\textbf{For researchers.}}  Significant research efforts are required to enable transparency, auditability, and accountability in the EdTech ecosystem.
Researchers could develop non-intrusive tools to record and analyze data flows for oversight and auditing purposes, especially post-acquisition. 
Privacy researchers could study techniques to establish private-by-design paradigms, such as differential privacy for aggregated data, minimal purpose-specific data collection, and verifiable data deletion protocols. 
Legal scholars could scrutinize existing laws to identify policy gaps that vendors exploit and create more comprehensive frameworks benefiting HEIs and student privacy. Researchers in business and management could investigate the impact of company acquisitions on data policies and handling practices, particularly for large vendors like Canvas. 
Finally, policy researchers could develop methods to ensure vendors comply with data protection standards before and after procurement, enabling HEIs to be confident that vendors are not misusing or mishandling their data throughout the procurement process.

\noindent\textit{\textbf{For HEIs.}} We recommend HEIs increase inter-HEI collaborations to exchange knowledge about best practices and experiences with vendors, which will help strengthen their data security and privacy posture. 
Forming coalitions will enable HEIs to be more confident in their vendor agreements and enforce better data protection practices. 
For example, by being upfront and asking for clarifications on vague clauses, supply chains (including third-party vendors, sub-vendors, shadow IT, etc.), and the off-boarding process.
Additionally, involving cybersecurity and policy can help HEIs design rigorous vendor assessment processes, negotiate clear data use agreements, and conduct regular audits and compliance checks. 
HEIs could also adopt data minimization practices, advocate for stronger regulations, and leverage more third-party security assessments.

We end with optimism noting that some HEIs are forming data governance boards, we hope that \sp researchers will collaborate to identify appropriate privacy-enhancing tools, as well as research and develop new tools that meet the specific needs of education processes. 

%% file: appendix.tex
\clearpage
\section*{Appendix}
\appendix

\section{Interview Questions}
\label{interview_questions}

\subsection{Policy Posture}
\begin{enumerate}
    \item Can you give me a glimpse into the considerations of your institute’s privacy policy? And talk to me about the latest changes made to the policy?

\begin{enumerate}
        \item What federal and state level data policies are used as the base policies?
        \item What special considerations are in place at your university?
\end{enumerate}

    \item What federal and state level data policies are used as guidelines when creating privacy policies and revisions?
Does the university policy for EdTech vendors contain additional data protection measures beyond fed/state level policies? How were they determined? Who was involved?

    \item Follow up to previous ques: Is your university data policy completely based on Fed/state laws, or are there enhancements to those? How were those enhancements/amendments made and are kept updated?
    \item What are the challenges you faced in developing data privacy policy? (Complications, cost of changing policies, impacts, etc.) 
    \item How often are your institution’s privacy policies revised and revisited?

\begin{enumerate}
        \item Who is involved with revisions?
        \item What causes a revision to take place?
\end{enumerate}

    \item As HEIs are increasingly offloading computing services to third party for-profit organizations, is there any back-up plan for when this becomes unsustainable? For example, Google Drive initially offered unlimited storage for Faculty members, but later started charging fees if the amount of data exceeds some threshold. Many had no way but to pay the fee since they become reliant on the service. Future price hikes can be arbitrarily high, once all doors to stop using such services get closed. Is there any plan to cope with such situations?
    \item Related to the previous question, how does academic freedom get impacted as we are relying more on technology? For example, services like zoom and github bans certain countries, zoom even prevented some academic meetings from taking place on the platform. How prepared is the university administration to handle such situations? 
\end{enumerate}

\subsection{Acquisition Phase}

\begin{enumerate}
    \item What is the whole process of acquiring an EdTech?

\begin{enumerate}
        \item What triggers the starting of the process to acquire/license a tech? How the candidate set of tech/vendors are selected? What factors are considered in that selection process?
        \item What is the typical timeline of a university acquiring an Edtech, from first wanting a new technology to implementation?
What takes the most amount of time? (testing/paperwork/price negotiation/etc)

        \item What type of meetings take place? Who is involved?

    \begin{enumerate}
            \item Who is involved in the process when choosing to acquire a new EdTech?
            \item Is there any way to incorporate inputs from faculty and/or students? How can the end users raise concerns about tools the institute rolls out? 
    \end{enumerate}

        \item What type of tests and/or trials are done? 
\end{enumerate}

    \item What factors dictate what EdTech the school purchases/license? 

\begin{enumerate}
        \item How do vendors attract universities to acquire their product?
        \item Which of those factors matter the most?
        \item What restricts an institution from getting an EdTech?
How many different vendors does the institution buy from?

        \item What is the oldest EdTech that your institution has retained?
How often are EdTechs removed from the institution’s list/How often is the EdTech inventory audited? How are EdTechs with overlapping functionalities and services brought in/handled?

\end{enumerate}

    \item Can you remember a vendor that you worked directly with and explain how the terms of service came to fruition?

\begin{enumerate}
        \item What adjustments are likely when the EdTech's privacy policy does not align with the institute's?
        \item How are price, service contract, data storage and usage policy, etc.  negotiated? Have you been actively or passively involved in negotiation with a vendor?
\end{enumerate}

    \item During negotiations, is there a specified person that reads what data a vendor collects? What type of data do they collect from users?

\begin{enumerate}
        \item Are there any data collections that make a university unwilling to acquire an EdTech? What are they?
        \item Was there any data breach or other security incidents involving your institution?

    \begin{enumerate}
            \item How was that incident handled by the institution and associated vendors?
    \end{enumerate}

\end{enumerate}
    \item  Do the members in the edtech acquisition/evaluation team review recent relevant literature? What measures are in place to incorporate current research into such decision making?
    \item What are the approaches that XYZ department takes to stay current with the privacy policies, compliances, as well as the security technologies used by the EdTech vendors?

    \item What are the barriers to augmenting EdTech with privacy-protective mechanisms? If [your institution] wants to augment Canvas with such mechanisms, what needs to be done?
    \item What, according to you, would be an ideal acquisition of EdTech?
\end{enumerate}

\subsection{Post-Acquisition Phase}
\begin{enumerate}
    \item How does the institute negotiate data ownership and the subsequent use of said data?

        \item Who holds responsibility for preventing abuse and the protection of the collected data?
        \item How are contracts used to avoid the abuse of data?

    \item Even if the data is owned/maintained by the university, can the service providers use that data for product improvement, marketing, or any other purpose? How do you make sure that the data is not used inappropriately? What happens to their access to the data when the service is discontinued?
    \item When implemented into the university or during testing, were there any anomalies during operation of an EdTech?
How do you handle those?

    \item How does the institution handle any data that could be shared between vendors?
        \item Do certain vendors get more access than others?
What allows them more access?

        \item How are contracts between vendors handled?
        \item What specific measures are taken to prevent malware or ransomware?
Did they observe any malware or ransomware spread, and how was that handled?

    \item What happens when devices/vendors are discontinued? When discarding devices, is there any standard procedure to make sure that the data is wiped out? When discontinuing any service, can the vendor continue storing/using the collected data?
    \item How does your institution test whether the technology behaves in accordance with the contract regarding data collection, (secondary) use, and sharing with business partners? What can be done if you find out that they violated the policy/contract? EdTech goes against the university's privacy policy?
    \item In case of contract violation, can you end the contract immediately? If the vendor collected or used data in ways that are not allowed by the contract, what can be done?
    \item What would you consider as the ideal setup for data handling, storage, and management within your organization and the overall EdTech ecosystem? This could include aspects like data security, accessibility, scalability, and efficient retrieval.
\end{enumerate}
 
\input{table_participants}

\section{Supplementary Study Materials}
\label{supplememtary_materials}
The codebook for this study, containing theme names, category names, category descriptions, and open codes, along with the recruitment email can be accessed here: \url{https://osf.io/84y2h/?view_only=1f0f36c5673a4ebdb78e49e22f319b66}

%% file: table_participants.tex
\begin{table*}[tb]
        \small
	\renewcommand{\arraystretch}{1.4} 
	\caption{Participant Demographics}
	\label{tab:participants-info}
        \centering
	\begin{tabular}{|p{1cm} | p{4.5cm} |p{4.5cm} | p{1cm} | p{1.5cm} | p{2cm}|}
		\hline
		\textbf{ID} & \textbf{Current Job Title} & \textbf{Role in EdTech Acquisition (Years in current Role)} & \textbf{Years of Exp. in EdTech}& \textbf{EdTech Handled} & \textbf{Region}  \\
		\hline
  
P1 & Chief Technology Officer (CTO) & Oversees security review  (4) & 23& \textasciitilde 100 & West Coast \\  
            P2 &Chief Information Officer and VP for Information Technology & Oversees security audits (4) & 15&200+&East Coast\\ 
 P3 &Deputy Chief Information Officer  & Oversees security audits (2)&23&-&Mid-west\\ 
 
 P4  & Chief Information Security Officer &Oversees security reviews and audits  (3)& 3&200+&South\\ 

P5 & Chief Information Security Officer  &Oversees procurement of EdTech handling sensitive data (5)&10&-&Mid-west\\ 

P6 & Deputy Chief Information Officer &Implements privacy controls in contracts (4)&4&200+&South\\ 

P7 & Senior Director of Learning &Oversees procurement (7)&7&200+&South\\ 

P8 &Senior Director of Learning Experience and IT Services&Oversees procurement (4)&4&200+&South\\ 
P9 & Executive Director of Data Analysis &Handles on premise technology data storage (5)&22&200+&South\\ 
P10 & Enterprise Partner &Negotiations with vendors (13) &13&200+&South\\ 
P11 &Deputy Chief Information Officer  &Oversees security reviews and purchasing committees (9) &25&\textasciitilde100&Mid-west\\ 
P12 &Chief Information Security Officer &Oversees security reviews and data handling (2) &12&200+&North-east\\ 
P13 &IT Support &Handles Edtech post-acquisition (5)&5&\textasciitilde100&South\\

            \hline
	\end{tabular}
\end{table*}